# All ℓ-state eigensolutions of the non-relativistic Schrodinger equation with general molecular oscillator

D. Yabwa [1], E.S. Eyube [2], A. D. Abu Ubaida[3] and V. Targema[4]

[1,4] Department of Physics, Faculty of Science, Taraba State University, Jalingo, Nigeria

[2,3] Department of Physics, School of Physical Sciences, Modibbo Adama University of Technology, Yola, Nigeria

[1] Corresponding Author: Email: yabwa.dlama@tsuniversity.edu.ng Tel: +2348133275536

**Abstract:** In this study, we employ exact quantization rule (EQR) to derive the analytical approximate ℓ-wave solutions of the Schrodinger equation with the general molecular oscillator(GMO). The energy eigenvalues equation and the corresponding wave-functions have been obtained explicitly. Improved Pekeris-type approximation Schemes have been used to deal with the orbital centrifugal term. We have deduced expressions for the bound-state energy eigenvalues of the Morse and shifted Deng-fan oscillator as special cases of the GMO, and computed their resulting bound-state energy eigenvalues for $H_2$, CO, HCl and LiH diatomic molecules. Our results are in good agreement with other results in the literature.

**1. Introduction**

Research in quantum mechanics has shown that the exact solution of wave equations is fundamental to understanding the molecular spectrum of a diatomic molecules in a non-relativistic quantum system [1-2]. The exact analytical solutions of the wave equations are possible for only few potentials of physical interest such as, the Coulombic potential and harmonic oscillator potential [3-4] for all quantum states $n\ell$ where n is the vibrational quantum number and ℓ is the rotational quantum number [5], similarly, few other potential energy functions such as Eckart, Hulthén and Morse potentials give exact solution only for the s-wave (ℓ = 0) state [6-7]. Most of the known potential energy functions have no exact solutions with the Schrödinger equation for all values of *n* and ℓ, for such potentials, approximate solutions (numerical or analytical) can be used in place of exact solutions [8]. In order to obtain approximate analytical solution, a very suitable approximation scheme [9-11] must be applied on the spin-orbit term of the effective potential, having applied the approximation model on the centrifugal term, a solution method must be adopted to solve the resulting equation. Over the last few decades a variety of analytical solution methods such as: ansatz method [12], generalized pseudospectral method [13], proper quantization rule [14-15], path integral approach [16], Laplace transform approach [4], Nikiforov-Uvarov method [17-18], asymptotic iteration method [19] have been developed to solve the wave equations exactly for physically interesting potentials. Recently an alternative method for solving radial Schrödinger equation for a given potential energy function was developed by Ma and Xu [20] which is called exact quantization rule (EQR) to find the exact energy eigenvalues for Coulombic and harmonic oscillator potentials.  By employing this method Qiang *et al*. have solved the arbitrary ℓ-state approximate solutions of the Hulthén potential [21]. Moreover, Ikhdair and Sever applied the exact quantization rule to the Kratzer-type oscillator and obtained the energy eigenvalues for some diatomic molecule [1], their results show that the EQR is a powerful tool in finding the eigenvalues of all solvable quantum oscillators especially for large values of n and ℓ. In describing a suitable molecular potential for diatomic molecule, researchers have developed and used various forms of potential energy functions, Jia *et al*. have shown that improved Tietz potential is equivalent with the Wei potential for diatomic molecules [22], Wang *et al*. have defined an improved form of Manning-Rosen and Schiöberg potentials and proved that the Deng-Fan and the improved forms of Manning-Rosen and Schiöberg potentials they are same solvable potentials [23]. The answer to which potential is more successful in explaining experimental data, will depends on the examined molecule, for these reasons it can be said that none of these potentials are sufficiently successful on their own, thus in order to overcome this problem, Yanar *et al.* proposed the general molecular oscillator, to model all diatomic molecule, they used the GMO to model diatomic molecules and successfully computed the vibrational energies of some molecules which were consistent with RKR data, they also showed that the general molecular potential can be reduced to Rosen-Morse potential which gave outstanding results in explaining the vibration of nitrogen in ammonia molecule [24]. In previous study, Ikot *et al.* obtained the energy spectral of general molecular oscillator and used their results to explore the thermodynamic properties of this potential [25]. Motivated by the successes in the application of EQR as a solution method of the Schrodinger equation, in this paper, we are encouraged to solve the radial Schrodinger equation with the GMO, to obtain closed form expressions for bound state energy eigenvalues and radial wave-functions, to test the accuracy of our results we will deduce closed form expressions for the bound state energy eigenvalues of Morse and shifted Deng-fan oscillator which are special cases of the GMO and compare our results with the existing literature. To the best of our knowledge this approach of EQR on GMO was not done before.

**2. Review of the concepts of exact quantization rule**

Here we present a brief introduction of the exact quantization rule as proposed by Ma and Xu [20]. The EQR has been proposed to solve the one-dimensional Schrödinger equation:

$$\psi''_{n\ell}(x) + k^2_{n\ell}(x)\psi_{n\ell}(x) = 0 \tag{1}$$

where

$$k_{n\ell}(x) = \sqrt{\frac{2\mu}{\hbar^2}\{E_{n\ell} - V_{eff}(x)\}} \tag{2}$$

is the momentum of the system, µ as the mass, $E_{n\ell}$ is the energy eigenvalue, $V_{eff}(x)$ is the effective potential energy function which is a piecewise continuous real function of $x$ and $\psi_{n\ell}(x)$ is the wave function. If we define the phase angle $\varphi_{n\ell}(x)$ given as:

$$\varphi_{n\ell}(x) = \psi'_{n\ell}(x)/\psi_{n\ell}(x) \tag{3}$$

Eq. (1) can be reduced into Riccati differential equation given by:

$$\varphi'_{n\ell}(x) + \frac{2\mu}{\hbar^2}\{E_{n\ell} - V_{eff}(x)\} + \varphi^2_{n\ell}(x) = 0 \tag{4}$$

where the phase angle, $\varphi_{n\ell}(x)$ represent the logarithmic derivative of the wave function, given as:

$$\varphi_{n\ell}(x) = \psi'_{n\ell}(x)/\psi_{n\ell}(x) \tag{5}$$

Due to Sturm-Liouville theorem, $\varphi_{n\ell}(x)$ decreases monotonically with respect to $x$ between two turning points determined by the equation $E_{n\ell} \geq V_{eff}(x)$. Specifically, as $x$ increases across a node of the wavefunction $\psi_{n\ell}(x)$, where $E_{n\ell} \geq V_{eff}(x)$, $\varphi_{n\ell}(x)$ decreases to $-\infty$ and jumps to $+\infty$ and then decreases again. By carefully studying the 1-D Schrödinger equation, Ma and Xu [20], proposed an EQR given by:

$$\int_{x_{nA}}^{x_{nB}} k_{n\ell}(x)dx = N\pi + \int_{x_{nA}}^{x_{nB}} \varphi_{n\ell}(x)\left[\frac{dk_{n\ell}(x)}{dx}\right]\left[\frac{d\varphi_{n\ell}(x)}{dx}\right]^{-1} dx \qquad (6)$$

where $x_{nA}$ and $x_{nB}$ are two turning points determined by letting $E_{n\ell} = V_{eff}(x)$ and $x_{nA} < x_{nB}$. N is the number of nodes of $\varphi_{n\ell}(x)$ in the neighborhood of $E_{n\ell} \geq V_{eff}(x)$ and it is larger by one than the number of nodes n of the wavefunction $\psi_{n\ell}(x)$, clearly, N = n + 1. The first term, N$\pi$, is the contribution from the nodes of the wave function, and the second term is referred to as the quantum correction. The quantum correction is independent of the number of nodes for the exactly solvable systems [20], therefore, it can be evaluated for the ground state (n = 0), the second term in Eq. (6) can be simply written as:

$$Q_c = \int_{x_{nA}}^{x_{nB}} \varphi_{n\ell}(x)\left[\frac{dk_{n\ell}(x)}{dx}\right]\left[\frac{d\varphi_{n\ell}(x)}{dx}\right]^{-1} dx \equiv \int_{x_{0A}}^{x_{0B}} \varphi_{0\ell}(x)\left[\frac{dk_{0\ell}(x)}{dx}\right]\left[\frac{d\varphi_{0\ell}(x)}{dx}\right]^{-1} dx \qquad (7)$$

where $Q_c$ is the quantum correction term. In three dimensional spherical coordinates, the EQR is given by:

$$\int_{r_{nA}}^{r_{nB}} k_{n\ell}(r)dr = N\pi + \int_{r_{nA}}^{r_{nB}} \varphi_{0\ell}(r)\left[\frac{dk_{0\ell}(r)}{dr}\right]\left[\frac{d\varphi_{0\ell}(r)}{dr}\right]^{-1} dr \qquad (8)$$

In simplified form, Eq. (7) can be expressed as:

$$I = N\pi + Q_c \qquad (9)$$

where the momentum integral is given by:

$$I = \int_{r_{nA}}^{r_{nB}} k_{n\ell}(r)dr \qquad (10)$$

and the quantum correction is:

$$Q_c = \int_{r_{nA}}^{r_{nB}} \phi_{0\ell}(r)\left[\frac{dk_{0\ell}(r)}{dr}\right]\left[\frac{d\phi_{0\ell}(r)}{dr}\right]^{-1} dr \qquad (11)$$

The radial Schrödinger equation in three-dimensional spherical coordinates [5] is written as:

$$\frac{d^2\psi_{n\ell}(r)}{dr^2} + \frac{2\mu}{\hbar^2}\{E_{n\ell} - V_{eff}(r)\}\psi_{n\ell}(r) = 0 \qquad (12)$$

$\psi_{n\ell}(r)$ is the radial wave function, r is the internuclear separation and $V_{eff}(r)$ is the effective potential defined in terms of a spherically symmetric potential $V(r)$ and a parameter L $=\ell(\ell+1)$ by:

$$V_{eff}(r) = V(r) + \frac{L\hbar^2}{2\mu r^2} \qquad (13)$$

### 3. The energy eigenvalues and eigenfunctions of the generalized molecular oscillator

The general molecular oscillator can be expressed in the form [24]:

$$V(r) = De\left\{\frac{E + Fe^{-\delta(r-r_e)} + Ge^{-2\delta(r-r_e)}}{\left(1 - qe^{-\delta(r-r_e)}\right)^2}\right\} \qquad (14)$$

where E, F, G, δ are adjustable potential parameters, q is a dimensionless parameter and $r_e$, is the equilibrium bond length. The effective potential is given by:

$$V(u) = De\left\{\frac{E + Fe^{-\alpha u} + Ge^{-2\alpha u}}{\left(1 - qe^{-\alpha u}\right)^2}\right\} + \frac{L\hbar^2}{2\mu r^2} \qquad (15)$$

where $\alpha = \delta r_e$, $u = r/(r_e-1)$, $L = \ell(\ell+1)$ and ℏ is the reduced Planck's constant

For a given value of parameters: E, F, G and q, Eqs. (15) and (12) have solution only for the pure vibrational state (ℓ = 0). In order to obtain analytical solution for all states, an approximation scheme must be used on the spin-orbit or centrifugal term of Eq. (15), therefore, we invoke the Pekeris-type approximation model [28] given by:

$$\frac{1}{r^2} \approx \frac{1}{r_e^2}\left\{c_0 + \frac{c_1}{e^{\alpha u} - q} + \frac{c_2}{(e^{\alpha u} - q)^2}\right\} \qquad (16)$$

where $c_0$, $c_1$ and $c_2$ are constants given by:

$$c_0 = 1 - \frac{1}{\alpha}(1-q)(3+q) + \frac{3}{\alpha^2}(1-q)^2 \qquad (17)$$

$$c_1 = \frac{2}{\alpha}(1-q)^2(2+q) - \frac{6}{\alpha^2}(1-q)^3 \qquad (18)$$

$$c_2 = -\frac{1}{\alpha}(1-q)^3(1+q) + \frac{3}{\alpha^2}(1-q)^4 \qquad (19)$$

However, by taking $c_0$, $c_1$, and $c_2$ as $4d_0 r_e^2 \delta^2$, $r_e^2 \delta^2$ and $r_e^2 \delta^2$ respectively where $d_0$ has the value ⅓, the Pekeris-type approximation reduces to the improved Green and Aldrich approximation model [27], we shall use these approximation schemes to test the accuracy of our results where appropriate. By substituting Eq. (16) in (15), the effective potential is then:

$$V_{eff}(u) = De\left\{\frac{E + Fe^{-\alpha u} + Ge^{-2\alpha u}}{\left(1 - qe^{-\alpha u}\right)^2}\right\} + \frac{L\hbar^2}{2\mu r_e^2}\left\{c_0 + \frac{c_1 e^{-\alpha u}}{1 - qe^{-\alpha u}} + \frac{c_2 e^{-2\alpha u}}{\left(1 - qe^{-\alpha u}\right)^2}\right\}$$

(20)

using the following transformation on Eq. (20)

$$x = \frac{e^{-\alpha u}}{1 - qe^{-\alpha u}} \qquad (21)$$

gives

$$V_{eff}(x) = \left\{ \frac{L\hbar^2 c_2}{2\mu r_e^2} + D_e \left( q^2 E + q F + G \right) \right\} x^2 + \left\{ \frac{L\hbar^2 c_1}{2\mu r_e^2} + D_e \left( 2q E + F \right) \right\} x + \frac{L\hbar^2 c_0}{2\mu r_e^2} + D_e E$$

(22)

For brevity, Eq. (22) can be written in the more compact form as:

$$V_{eff}(x) = A x^2 + B x + C \tag{23}$$

with the constants A, B and C given by:

$$A = \frac{\hbar^2 \varepsilon_1^2}{2\mu r_e^2} \tag{24}$$

$$B = \frac{\hbar^2 \varepsilon_2^2}{2\mu r_e^2} \tag{25}$$

$$C = \frac{\hbar^2 \varepsilon_3^2}{2\mu r_e^2} \tag{26}$$

where

$$\varepsilon_1^2 = L c_2 + \frac{2\mu r_e^2 D_e}{\hbar^2} \left( q^2 E + q F + G \right) \tag{27}$$

$$\varepsilon_2^2 = L c_1 + \frac{2\mu r_e^2 D_e}{\hbar^2} \left( 2q E + F \right) \tag{28}$$

$$\varepsilon_3^2 = L c_0 + \frac{2\mu r_e^2 D_e E}{\hbar^2} \tag{29}$$

Following Ma and Xu (2005), we can determine the turning points $x_{nA}$ and $x_{nB} (> x_{nA})$ by imposing the condition:

$$V_{eff}(x) = E_{n\ell} \tag{30}$$

Substituting Eq. (23) in (30), we find the sum $x_{nA} + x_{nB}$ and the product $x_{nA} x_{nB}$ as given by:

$$x_{nA} + x_{nB} = -\frac{B}{A} \equiv -\frac{\varepsilon_2^2}{\varepsilon_1^2} \tag{31}$$

$$x_{nA} x_{nB} = \frac{C - E_{n\ell}}{A} \tag{32}$$

In what will be required in the evaluation of quantum correction, the ground state equivalents of Eqs. (31) and (32) are required, therefore putting n = 0 in Eqs. (31) and (32), we have:

$$x_{0A} + x_{0B} = -\frac{\varepsilon_2^2}{\varepsilon_1^2} \tag{33}$$

$$x_{0A} x_{0B} = \frac{C - E_{0\ell}}{A} \tag{34}$$

Clearly, the sum $x_{nA} + x_{nB}$ is independent of the vibrational and rotational quantum numbers.

The expression for the momentum of the system is given by:

$$K_{n\ell}^2 = \frac{2\mu}{\hbar^2}\left(E_{n\ell} - Ax^2 - Bx - C\right) \tag{35}$$

Expressing Eq. (35) in terms of the turning points, this results to:

$$k_{n\ell}(x) = \frac{\varepsilon_1}{r_e}\sqrt{(x - x_{nA})(x_{nB} - x)} \tag{36}$$

And the derivative of Eq. (36) is given by:

$$k'_{n\ell}(x) = \frac{\varepsilon_1}{r_e}\frac{x + \varepsilon_2^2/2\varepsilon_1^2}{\sqrt{(x - x_{nA})(x_{nB} - x)}} \tag{37}$$

Evaluating the derivative of the momentum for the ground state, this gives:

$$k'_{0\ell}(x) = \frac{\varepsilon_1}{r_e}\frac{x + \varepsilon_2^2/2\varepsilon_1^2}{\sqrt{(x - x_{0A})(x_{0B} - x)}} \tag{38}$$

expressing the Riccati equation given by Eq. (4) in terms of variable x, we have that:

$$-\frac{\alpha}{r_e}\left(x + qx^2\right)\varphi'_{n\ell}(x) + \varphi_{n\ell}^2(x) + \frac{2\mu}{\hbar^2}\left(E_{n\ell} - Ax^2 - Bx - C\right) = 0 \tag{39}$$

which for the ground state gives

$$-\frac{\alpha}{r_e}\left(x + qx^2\right)\varphi'_{0\ell}(x) + \varphi_{0\ell}^2(x) + \frac{2\mu}{\hbar^2}\left(E_{0\ell} - Ax^2 - Bx - C\right) = 0 \tag{40}$$

For a trial solution of Eq. (40), choose:

$$\varphi_{0\ell}(x) = -a_1 x + a_2 \tag{41}$$

where $a_1$ and $a_2$ are constants. Substituting Eq. (41) in Eq. (40) gives:

$$\frac{a_1 \alpha}{r_e}\left(x + qx^2\right) + a_1^2 x^2 - 2a_1 a_2 x + a_2^2 + \frac{2\mu}{\hbar^2}\left(E_{0\ell} - Ax^2 - Bx - C\right) = 0 \tag{42}$$

From Eq. (42) we have:

$$\left(a_1^2 + \frac{a_1 \alpha q}{r_e} - \frac{2\mu A}{\hbar^2}\right)x^2 + \left(\frac{a_1 \alpha}{r_e} - 2a_1 a_2 - \frac{2\mu B}{\hbar^2}\right)x + \left(a_2^2 - \frac{2\mu}{\hbar^2}(E_{0\ell} - C)\right) = 0$$

(43)

equate the corresponding coefficients of $x^2$, x and $x^0$ respectively on both sides of the Eq. (43) we have the following equations:

$$\frac{a_1 \alpha q}{r_e} + a_1^2 = \frac{2\mu}{\hbar^2} A \equiv \frac{\varepsilon_1^2}{r_e^2} \tag{44}$$

$$\frac{a_1 \alpha}{r_e} - 2a_1 a_2 = \frac{2\mu}{\hbar^2} B \equiv \frac{\varepsilon_2^2}{r_e^2} \tag{45}$$

$$a_2^2 = \frac{2\mu}{\hbar^2}(C - E_{0\ell}) \tag{46}$$

$$a_1 = -\frac{\alpha q}{r_e}\left\{\frac{1}{2} + \left(\frac{1}{4} + \frac{\varepsilon_1^2}{\alpha^2 q^2}\right)^{\frac{1}{2}}\right\} \tag{47}$$

we have

$$a_1 = -\frac{\alpha q \sigma}{r_e} \tag{48}$$

where

$$\sigma = \frac{1}{2} + \left(\frac{1}{4} + \frac{\varepsilon_1^2}{\alpha^2 q^2}\right)^{\frac{1}{2}} \tag{49}$$

from Eq. (44) and (45), we can get our constants and thus find our trial solution
Eq. (11) gives the quantum correction in terms of the variable x as:

$$Q_c = -\frac{r_e}{\alpha}\int_{x_{0A}}^{x_{nB}} \frac{\varphi_{0\ell}(x)}{\varphi'_{0\ell}(x)} k'_{0\ell}(x) \frac{dx}{x(1+qx)} \tag{50}$$

substituting Eq. (41) and (37) and taking n = 0 in Eq. (50), we have

$$Q_c = \frac{\varepsilon_1}{\alpha}\int_{x_{0A}}^{x_{nB}} \frac{(x - a_2/a_1)(x + \varepsilon_2^2/2\varepsilon_1^2)}{x(1+qx)} \frac{dx}{\sqrt{(x-x_{0A})(x_{0B}-x)}} \tag{51}$$

the above integral can be evaluated by standard integral eq. (A1) [28], which gives:

$$Q_c = \frac{\pi \varepsilon_1}{\alpha}\left\{\frac{1}{q} - \frac{a_2 \varepsilon_2^2}{2 a_1 \varepsilon_1^2}\frac{\varepsilon_1}{a_2 r_e} + \frac{(a_1 + q a_2)(q \varepsilon_2^2 - 2\varepsilon_1^2)}{2 a_1 q \varepsilon_1^2}\frac{\varepsilon_1}{(a_1 + a_2 q)r_e}\right\} \tag{52}$$

Eq. (52) simplifies to:

$$Q_c = \frac{\pi \varepsilon_1}{\alpha}\left(\frac{1}{q} + \frac{\varepsilon_1}{\alpha \sigma q^2}\right) \tag{53}$$

Next, we evaluate the momentum integral on the left hand side of Eq. (8) and applying the transformation given by Eq. (21), we have:

$$I = -\frac{\varepsilon_1}{\alpha}\int_{x_{nA}}^{x_{nB}} \frac{\sqrt{(x-x_{nA})(x_{nB}-x)}dx}{x(1+qx)} \tag{54}$$

The momentum integral of Eq. (36) was used in arriving at Eq. (54). In order to evaluate the definite integral in (54), we employ the following standard integral Eq. (A2) [28], therefore, Eq. (54) gives:

$$I = -\frac{\pi \varepsilon_1}{\alpha}\left\{\frac{\sqrt{(q x_{nA} + 1)(q x_{nB} + 1)}}{q} - \frac{1}{q} - \sqrt{x_{nA} x_{nB}}\right\} \tag{55}$$

It follows that by using Eqs. (32), (24), (26) and (55), the energy eigenvalues of the GMO is given by:

$$E_{n\ell} = \frac{\hbar^2 \varepsilon_3^2}{2\mu r_e^2} - \frac{\hbar^2 \alpha^2}{2\mu r_e^2}\left\{\frac{\varepsilon_1^2 - q\varepsilon_2^2}{2\alpha^2 q^2 (n+\sigma)} - \frac{n+\sigma}{2}\right\}^2 \tag{56}$$

Using Eqs. (27) (28), (29) and (49) to eliminate $\varepsilon_1, \varepsilon_2, \varepsilon_3, \alpha$ and $\sigma$ in Eq. (56) we have:

$$E_{n\ell} = D_e E + \frac{L\hbar^2 c_0}{2\mu r_e^2} - \frac{\hbar^2 \delta^2}{2\mu}\left\{\frac{\frac{L}{\delta^2 q^2 r_e^2}(c_2 - qc_1) - \frac{2\mu D_e}{\delta^2 \hbar^2 q^2}(q^2 E - G)}{2n+1+\sqrt{1+\frac{4Lc_2}{\delta^2 q^2 r_e^2} + \frac{8\mu D_e}{\delta^2 \hbar^2 q^2}(q^2 E + qF + G)}} - \frac{2n+1+\sqrt{1+\frac{4Lc_2}{\delta^2 q^2 r_e^2} + \frac{8\mu D_e}{\delta^2 \hbar^2 q^2}(q^2 E + qF + G)}}{4}\right\}^2 \tag{57}$$

To obtain the energy eigenfunctions of the general molecular oscillator, we need to solve the Riccati equation given by (39), this will give the solution for the phase angle $\varphi_{n\ell}(x)$, therefore, by using the definition of the phase angle, the wave function, $\psi_{n\ell}(x)$ can be recovered. Using the following transformation equation:

$$z = 1 + qx \tag{58}$$

and the definition of the phase angle, Eq. (39) gives, with slight simplification:

$$z(1-z)\psi''_{n\ell}(z) + (1-2z)\psi'_{n\ell}(z) + \left\{\varepsilon_4 - \frac{\varepsilon_4 - \varepsilon_5 + \varepsilon_6}{z} - \frac{\varepsilon_6}{1-z}\right\}\psi_{n\ell}(z) = 0 \tag{59}$$

where

$$\varepsilon_4 = \frac{2\mu r_e^2 A}{\alpha^2 q^2 \hbar^2} \tag{60}$$

$$\varepsilon_5 = \frac{2\mu r_e^2 B}{\alpha^2 q \hbar^2} \tag{61}$$

$$\varepsilon_6 = \frac{2\mu r_e^2}{\alpha^2 \hbar^2}(C - E_{n\ell}) \tag{62}$$

Following [2] Eq. (59) has solution of the form:

$$\psi_{n\ell}(z) = N_{n\ell} z^\varsigma (1-z)^\xi \,_2F_1(-n, n+2\varsigma+2\xi+1; z) \tag{63}$$

where $_2F_1$ is the hypergeometric function and the constants $\varsigma$ and $\xi$ are subjected to the following constraints:

$$\varsigma = (\varepsilon_4 - \varepsilon_5 + \varepsilon_6)^{\frac{1}{2}} \tag{64}$$

$$\xi = \varepsilon_6^{\frac{1}{2}} \tag{65}$$

**4. Discussion**

The Morse oscillator

If we let q = 0, E = 1, F = -2 and G = 1 in Eq. (14), the effective potential reduces to:

$$V_{eff\_M}(r) = D_e \left\{1 - e^{-\delta(r-r_e)}\right\}^2 + \frac{L\hbar^2}{2\mu r^2} \tag{66}$$

Eq. (66) can be recognized to be the effective Morse oscillator, in which case the constant δ is the Morse constant [13,29]. Therefore, by inserting the above parameters in Eq. (57), the resulting energy eigenvalues are expected to give the energy eigenvalues ($E_{n\ell\_M}$) of Morse oscillator, however substituting these values in Eq. (57) leads to $E_{n\ell\_M} \to \infty$, which is physically not acceptable because $V_{eff\_M}(r)$ is a finite potential energy function. Thus, in order to obtain a physically acceptable expression for $E_{n\ell\_M}$ we need to re-evaluate the quantum correction $Q_{cM}$ and the momentum integral $I_M$ respectively for the Morse oscillator. Note that Eqs. (44) and (51) gives, for q = 0:

$$a_{1M} = -\frac{\varepsilon_1}{r_e} \tag{67}$$

$$Q_{cM} = -\frac{\varepsilon_1}{\alpha} \int_{x_{0A}}^{x_{nB}} \left(x + \frac{B}{2A} - \frac{a_{1M}}{a_{2M}} - \frac{a_{1M}B}{2a_{2M}Ax}\right) \frac{dx}{\sqrt{(x-x_{0A})(x_{0B}-x)}} \tag{68}$$

using standard integrals (A3-A6 in the appendix) obtained from ref. [28], Eq. (68) simplifies to:

$$Q_{cM} = -\frac{\pi}{2} \tag{69}$$

similarly, Eq. (54) gives for q = 0,

$$I_M = \frac{\varepsilon_1}{\alpha} \int_{x_{nA}}^{x_{nB}} \frac{\sqrt{(x-x_{nA})(x_{nB}-x)}dx}{x} \tag{70}$$

application of the standard integral (A6) on Eq. (70) gives:

$$I_M = -\frac{\pi\varepsilon_1}{\alpha}\left(\frac{\varepsilon_2^2}{2\varepsilon_1^2} + \sqrt{x_{nA}x_{nB}}\right) \tag{71}$$

Substituting Eqs. (72) and (69) in the exact quantization rule given by Eq. (9), we have that:

$$E_{n\ell\_M} = D_e E + \frac{L\hbar^2 c_0}{2\mu r_e^2} - \frac{\hbar^2 \delta^2}{2\mu}\left(n + \tfrac{1}{2} + \frac{\frac{Lc_1}{2\delta r_e} + \frac{\mu r_e D_e F}{\delta \hbar^2}}{\sqrt{Lc_2 + \frac{2\mu r_e^2 D_e G}{\hbar^2}}}\right)^2 \tag{72}$$

**Table 1**

Spectroscopic data of the molecules used in the present work, [6,27]

| Molecule | $D_e$ (eV) | $r_e$ (Å) | $\mu$ (amu) | $\delta$ (Å$^{-1}$) |
|---|---|---|---|---|
| $H_2$ | 4.7446 | 0.7416 | 0.50391 | 1.9506 |
| LiH | 2.5152672118 | 1.5956 | 0.8801221 | 1.128 |
| HCl | 4.619030905 | 1.2746 | 0.9801045 | 1.8677 |
| CO | 11.2256 | 1.1283 | 6.8606719 | 2.2994 |

Using Eq. (72) with q = 0, E = 1, F = -2 and G = 1, we have computed the bound state energy eigenvalues of the GMO, the computed results for the four diatomic molecules ($H_2$, CO, HCl and LiH) are shown in Tables 2 and 3, to enable comparison with existing literature results we have included in the table, the bound state energy eigenvalues for these molecules corresponding to the Morse oscillator obtained by generalized pseudospectral method (GPS) and the Nikiforov-Uvarov (NU) method. Firstly, it will be observed that the expressions for the Morse oscillator used in [31] has to be shifted by +De to agree with the GMO given by Eq. (66), for the purpose of comparing results, our computed bound state energy eigenvalues of Eq. (72) must be stepped down by –De, the results are shown in the tables from these results it is evident that the present results obtained by exact quantization rule agrees totally with those obtained by other method in the literature for both the low and the high lying quantum states.

**Table 2**

Bound state energy eigenvalues (in eV) for $H_2$, CO, HCl and LiH along with literature results

| states | | $H_2$ | | | CO | | | HCl | | | LiH | | |
|---|---|---|---|---|---|---|---|---|---|---|---|---|---|
| n | ℓ | PR | [31] | [32] | PR | [31] | [32] | PR | [31] | [32] | PR | [31] | [32] |
| 0 | 0 | 4.47490947 | 4.47601 | 4.47601313 | 11.09153513 | 11.0915 | 11.0910588 | 4.43552555 | 4.43556 | 4.43556394 | 2.42884375 | 2.42886 | 2.42886321 |
|   | 1 | 4.46012683 | … | 4.4612285 | 11.09105856 | … | 11.0901057 | 4.43293905 | … | 4.43297753 | 2.42700259 | … | 2.4270221 |
|   | 2 | 4.43069467 | … | 4.4317998 | 11.09010545 | … | 11.0901057 | 4.42776761 | … | 4.4278063 | 2.42332280 | … | 2.42334244 |
|   | 5 | 4.25770100 | 4.17644 | 4.17644 | 11.08438729 | 11.0844 | … | 4.39678297 | 4.39682 | … | 2.40131480 | 2.40133 | … |
|   | 10 | 3.72057413 | 3.72194 | 3.72194 | 11.06533305 | 11.0653 | … | 4.29404332 | 4.29408 | … | 2.32881725 | 2.32884 | … |
| 1 | 0 | 3.95920056 | … | 3.96231534 | 10.82582150 | … | 10.8258221 | 4.07967309 | … | 4.07971006 | 2.26052923 | … | 2.26054805 |
|   | 1 | 3.94497382 | … | 3.94811647 | 10.82534901 | … | 10.8253496 | 4.07716358 | … | 4.07720144 | 2.25873596 | … | 2.25875559 |
|   | 2 | 3.91664728 | … | 3.91986423 | 10.82440404 | … | 10.8244047 | 4.07214612 | … | 4.07218579 | 2.25515190 | … | 2.25517324 |
|   | 5 | 3.75011926 | … | … | 10.81873477 | … | … | 4.04208487 | … | … | 2.23371686 | … | … |
|   | 10 | 3.23266412 | … | … | 10.79984348 | … | … | 3.94241788 | … | … | 2.16311235 | … | … |
| 2 | 0 | 3.47505452 | … | 3.47991882 | 10.56332935 | … | 10.5633303 | 3.73869830 | … | 3.73873384 | 2.09825793 | … | 2.09827611 |
|   | 1 | 3.46138368 | … | 3.46633875 | 10.56286093 | … | 10.5628619 | 3.73626578 | … | 3.73630382 | 2.09651254 | … | 2.09653304 |
|   | 2 | 3.43416275 | … | 3.43932836 | 10.56192412 | … | 10.5619252 | 3.73140229 | … | 3.73144539 | 2.09302421 | … | 2.0930495 |
|   | 5 | 3.27410039 | … | … | 10.55630373 | … | … | 3.70226444 | … | … | 2.07216213 | … | … |
|   | 10 | 2.77631698 | … | … | 10.53757539 | … | … | 3.60567011 | … | … | 2.00345066 | … | … |
| 5 | 0 | 2.21199364 | 2.22052 | … | 9.79518178 | 9.79518 | … | 2.80503994 | 2.80506 | … | 1.64770335 | 1.64771 | … |
|   | 1 | 2.19999049 | … | … | 9.79472558 | … | … | 2.80283841 | … | … | 1.64610162 | … | … |
|   | 2 | 2.17608642 | … | … | 9.79381321 | … | … | 2.79843683 | … | … | 1.64290045 | … | … |
|   | 5 | 2.03542102 | 2.04355 | … | 9.78833949 | 9.78833 | … | 2.77206916 | 2.77209 | … | 1.62375726 | 1.62377 | … |
|   | 10 | 1.59665279 | 1.60391 | … | 9.77009999 | 9.77009 | … | 2.68469281 | 2.68471 | … | 1.56072491 | 1.56074 | … |
| 7 | 0 | 1.52776741 | 1.53744 | … | 9.29919080 | 9.29918 | … | 2.25698939 | 2.25701 | … | 1.37754972 | 1.37756 | … |
|   | 1 | 1.51687606 | … | … | 9.29874275 | … | … | 2.25494184 | … | … | 1.37604376 | … | … |
|   | 2 | 1.49518322 | … | … | 9.29784668 | … | … | 2.25084820 | … | … | 1.37303404 | … | … |
|   | 5 | 1.36744913 | 1.37656 | … | 9.29247073 | 9.29246 | … | 2.22632733 | 2.22634 | … | 1.35503678 | 1.35505 | … |
|   | 10 | 0.96802436 | 0.97581 | … | 9.27455713 | 9.27455 | … | 2.14509630 | 2.14511 | … | 1.29579050 | 1.2958 | … |

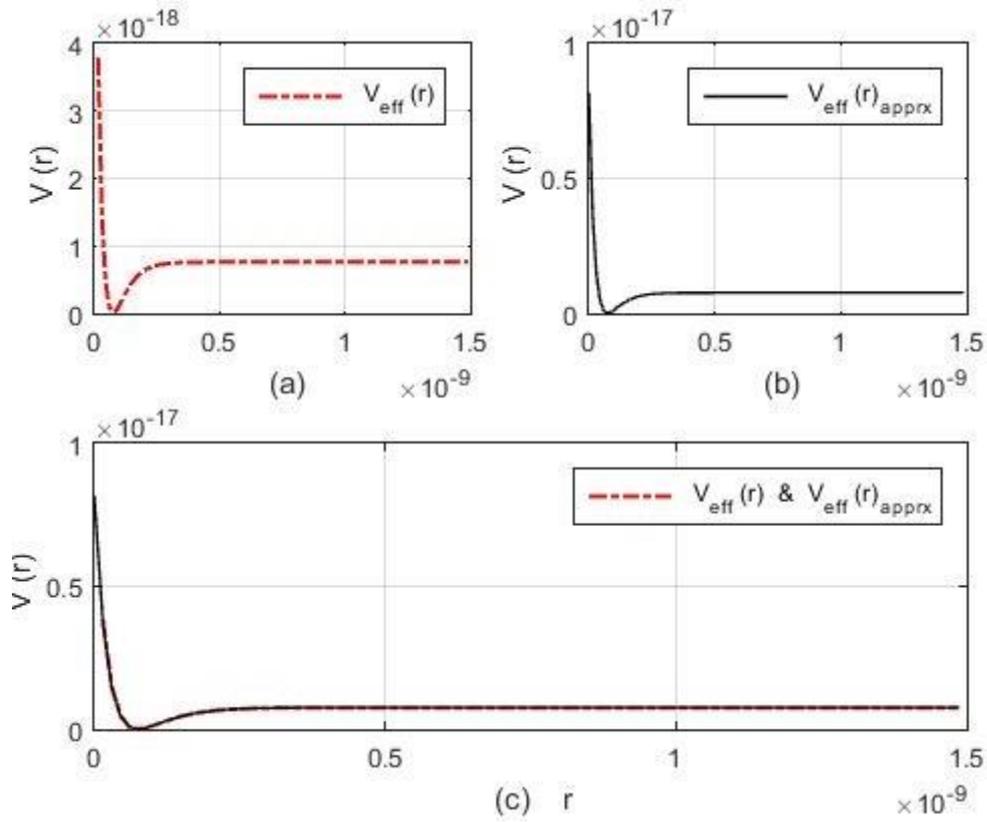

**Figure 1.** Plot of variation of effective general molecular potential Morse, V(r) in (a), and approximate effective general molecular potential Morse, V(r) in (b) with combined plot of (a) and (b) for $H_2$ in (c)

The generalized Morse oscillator

For values of the parameters E = 0, $F = 2(e^{-\delta r_e} - 1)$, $G = 1 - e^{-2\delta r_e}$ and $q = e^{-\delta r_e}$, the expression for the effective potential in Eq. (14) yields:

$$V_{eff\_SgM}(r) = D_e \left(1 - \frac{e^{\delta r_e} - 1}{e^{\delta r} - 1}\right)^2 - D_e + \frac{L\hbar^2}{2\mu r^2} \tag{73}$$

Eq. (73) is just the effective shifted generalized Morse oscillator, $V_{eff\_SgM}(r)$ [13]. It follows that upon substituting these values of the parameters in the expression for the energy eigenvalues of the general molecular oscillator given by Eq. (57), we have the energy eigenvalues for the shifted generalized Morse oscillator given as:

$$E_{n\ell\_SgM} = \frac{L\hbar^2 c_0}{2\mu r_e^2} - \frac{\hbar^2 \delta^2}{2\mu} \left\{ \frac{\frac{L}{\delta^2 r_e^2}(c_2 e^{\delta r_e} - c_1)e^{\delta r_e} + \frac{2\mu D_e}{\delta^2 \hbar^2}(e^{2\delta r_e} - 1)}{2n+1+\sqrt{1+\frac{4Lc_2 e^{2\delta r_e}}{\delta^2 r_e^2} + \frac{8\mu D_e}{\delta^2 \hbar^2}(e^{-\delta r_e}+1)^2}} - \frac{2n+1+\sqrt{1+\frac{4Lc_2 e^{2\delta r_e}}{\delta^2 r_e^2} + \frac{8\mu D_e}{\delta^2 \hbar^2}(e^{-\delta r_e}+1)^2}}{4} \right\}^2$$

(74)

To further test the accuracy our results we have employed the energy eigenvalues of the GMO of Eq. (74) to compute bound state energy eigenvalues for four diatomic molecules viz: $H_2$, CO, HCl and LiH the results of our computation are shown by the entries in Tables 4 and 5 and also the literature results which were obtained by alternative means of proper quantization rule (PQR) and asymptotic iteration method (AIM). The PQR results were available for $H_2$ and CO and upon comparison our present results is in near perfect agreement with the results for PQR. If we further compare the results EQR, PRQ and AIM, the results we obtained by EQR is almost indistinguishable with the other two methods except for few isolated cases of the $H_2$ molecule. The results obtained in this study is a clear demonstration of the efficacy of the EQR in obtaining bound state energy for diatomic molecules when considered for low and high quantum states.

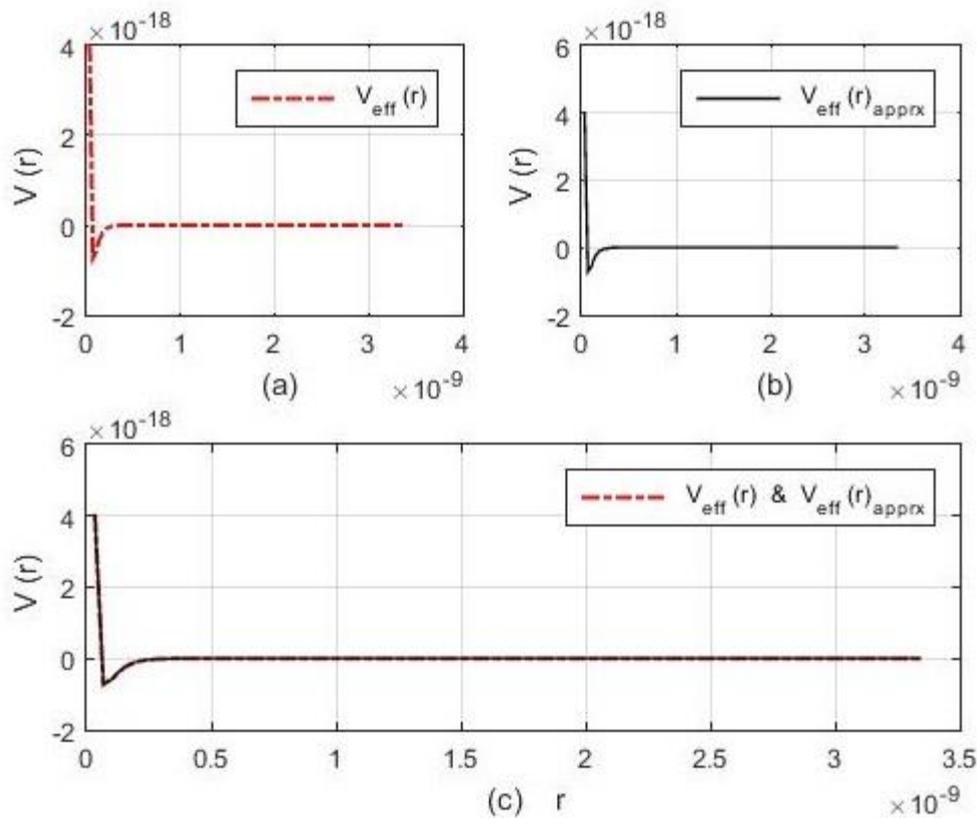

**Figure 2.** Plot of variation of effective general molecular potential V(r) in (a), and approximate effective general molecular potential V(r) in (b) with combined plot of (a) and (b) for HCl in (c).

**Table 3**

Bound state energy eigenvalues (in eV) for $H_2$, CO, HCl and LiH along with literature results

| states | | $H_2$ | | | CO | | | HCl | | LiH | |
|---|---|---|---|---|---|---|---|---|---|---|---|
| n | ℓ | PR | [30] | [28] | PR | [30] | [28] | PR | [30] | PR | [30] |
| 0 | 0 | 4.39383750 | 4.39461978 | 4.39461978 | 11.08075134 | 11.0807518 | 11.08075178 | 4.41704940 | 4.417077 | 2.41193393 | 2.41194905 |
|   | 1 | 4.37912480 | … | … | 11.08020383 | … | … | 4.41417939 | … | 2.41003189 | … |
|   | 2 | 4.34977450 | … | … | 11.07910881 | … | … | 4.40844043 | … | 2.40622945 | … |
|   | 5 | 4.17575583 | 4.17661805 | 4.17661316 | 11.07253907 | 11.0725399 | 11.07253746 | 4.37403672 | 4.37406578 | 2.38346076 | 2.38347625 |
|   | 10 | 3.62064027 | 3.62183842 | 3.62182049 | 11.05064413 | 11.0506458 | 11.05064208 | 4.25972896 | 4.25976195 | 2.30813095 | 2.30814747 |
| 1 | 0 | 3.74562763 | … | … | 10.79416640 | … | … | 4.02829079 | … | 2.21326520 | … |
|   | 1 | 3.73207676 | … | … | 10.79362281 | … | … | 4.02550106 | … | 2.21143254 | … |
|   | 2 | 3.70504523 | … | … | 10.79253563 | … | … | 4.01992265 | … | 2.20776881 | … |
|   | 5 | 3.54480384 | … | … | 10.78601289 | … | … | 3.98648179 | … | 2.18583121 | … |
|   | 10 | 3.03399921 | … | … | 10.76427465 | … | … | 3.87537764 | … | 2.11325695 | … |
| 2 | 0 | 3.16095471 | … | … | 10.51162718 | … | … | 3.65903222 | … | 2.02458767 | … |
|   | 1 | 3.14849738 | … | … | 10.51108749 | … | … | 3.65632173 | … | 2.02282281 | … |
|   | 2 | 3.12364843 | … | … | 10.51000812 | … | … | 3.65090178 | … | 2.01929464 | … |
|   | 5 | 2.97637761 | … | … | 10.50353225 | … | … | 3.61841118 | … | 1.99816919 | … |
|   | 10 | 2.50731885 | … | … | 10.48195021 | … | … | 3.51046879 | … | 1.92828778 | … |
| 5 | 0 | 1.75274100 | 1.75845157 | 1.75845157 | 9.68815923 | 9.68814619 | 9.688146187 | 2.66574180 | 2.66574902 | 1.51627313 | 1.51627729 |
|   | 1 | 1.74320803 | … | … | 9.68763118 | … | … | 2.66326292 | … | 1.51470268 | … |
|   | 2 | 1.72419619 | … | … | 9.68657510 | … | … | 2.65830616 | … | 1.51156322 | … |
|   | 5 | 1.61162673 | 1.61741062 | 1.61740572 | 9.68023898 | 9.68022628 | 9.680226284 | 2.62859335 | 2.62860119 | 1.49276676 | 1.49277143 |
|   | 10 | 1.25440404 | 1.26045164 | 1.26043371 | 9.65912268 | 9.659164 | 9.659110919 | 2.52989351 | 2.52990569 | 1.43060825 | 1.4306143 |
| 7 | 0 | 1.07146436 | 1.07763699 | 1.07763699 | 9.15918188 | 9.159164 | 9.159164003 | 2.09652455 | 2.0965248 | 1.22339249 | 1.22339354 |
|   | 1 | 1.06362899 | … | … | 9.15866154 | … | … | 2.09419521 | … | 1.22194465 | … |
|   | 2 | 1.04800595 | … | … | 9.15762087 | … | … | 2.08953750 | … | 1.21905034 | … |
|   | 5 | 0.95559189 | 0.96181478 | 0.96180989 | 9.15137718 | 9.15135966 | 9.151359661 | 2.06161809 | 2.06162002 | 1.20172274 | 1.20172434 |
|   | 10 | 0.66343343 | 0.66984407 | 0.66982613 | 9.13056901 | 9.13055243 | 9.130552425 | 1.96888561 | 1.96889204 | 1.14443549 | 1.14443859 |

## 5. Conclusion

In this research work, we have applied the ideas of exact quantization rule and ansatz solution method to obtain closed form expressions for the bound state rotational-vibrational eigensolutions of the GMO, cases of generalized Morse oscillator as well as generalized shifted Morse oscillator were considered. We have computed rotational-vibrational energies for four diatomic molecules *viz*: $H_2$, CO, HCl and LiH and compared our result with existing results in the literature. The results obtained in this work might be useful in areas of molecular physics, chemical physics, atomic physics and solid state physics.

**Appendix**

$$\int_{x_{nA}}^{x_{nB}} \frac{dx}{(P+Qx)\sqrt{(x-x_{nA})(x_{nB}-x)}} = \frac{\pi}{\sqrt{(P+Qx_{nB})(P+Qx_{nA})}} \quad (A1)$$

$$\int_{x_{nA}}^{x_{nB}} \frac{\sqrt{(x-x_{nA})(x_{nB}-x)}}{x(1+Qx)} dx = \pi \left\{ \frac{1}{Q} - \sqrt{x_{nA}x_{nB}} + \frac{\sqrt{(Qx_{nA}+1)(Qx_{nB}+1)}}{Q} \right\} \quad (A2)$$

$$\int_{x_{nA}}^{x_{nB}} \frac{x\,dx}{\sqrt{(x-x_{nA})(x_{nB}-x)}} = \frac{\pi}{2}(x_{nA}+x_{nB}) \quad (A3)$$

$$\int_{x_{nA}}^{x_{nB}} \frac{dx}{x\sqrt{(x-x_{nA})(x_{nB}-x)}} = \frac{\pi}{\sqrt{x_{nA}x_{nB}}} \quad (A4)$$

$$\int_{x_{nA}}^{x_{nB}} \frac{dx}{\sqrt{(x-x_{nA})(x_{nB}-x)}} = \pi \quad (A5)$$

$$\int_{x_{nA}}^{x_{nB}} \frac{\sqrt{(x-x_{nA})(x_{nB}-x)}}{x} dx = \pi \left\{ \tfrac{1}{2}(x_{nA}+x_{nB}) - \sqrt{x_{nA}x_{nB}} \right\} \quad (A6)$$